\documentclass[12pt]{article}
\usepackage[dvips]{epsfig}
\def \ln {{\rm ln}\, }

\newcommand{\be}{\begin{equation}}
\newcommand{\ee}{\end{equation}}
\newcommand{\bn}{\begin{eqnarray}}
\newcommand{\en}{\end{eqnarray}}

\newcommand{\nn}{\nonumber}

\def\xt{\tilde{x}}
\def\no{\noindent}
\def\bea{\begin{eqnarray}}
\def\eea{\end{eqnarray}}

\begin{document}
\title{\Large{\bf The Yang-Lee zeros of the 1D Blume-Capel model on connected
and non-connected rings }}
\bigskip
\author{{\bf Luis A. F. Almeida\thanks{e-mail:luisaug@feg.unesp.br}} \\
{\bf D. Dalmazi\thanks{e-mail:dalmazi@feg.unesp.br }}\\
{\small\it UNESP - Campus de Guaratinguet\'a - DFQ }\\
{\small\it Av. Dr. Ariberto P. da Cunha, 333, CEP 12516-410
  Guaratinguet\'a, SP, Brazil}}

\date{\today}
\maketitle
\begin{abstract}
We carry out a  numerical and analytic analysis of the Yang-Lee zeros of the 1D
Blume-Capel model with periodic boundary conditions and its generalization on Feynman
diagrams for which we include sums over all connected and non-connected rings for a
given number of spins. In both cases, for a specific range of the parameters, the
zeros originally on the unit circle are shown to departure from it as we increase the
temperature beyond some limit. The curve of zeros can bifurcate and become two
disjoint arcs as in the 2D case. We also show that in the thermodynamic limit the
zeros of both Blume-Capel models on the static (connected ring) and on the dynamical
(Feynman diagrams) lattice tend to overlap. In the special case of the 1D Ising model
on Feynman diagrams we can prove for arbitrary number of spins that the Yang-Lee zeros
must be on the unit circle. The proof is based on a property of the zeros of Legendre
Polynomials.
 \vfill
\noindent {\it PACS-No.}: 05.50.+q; 05.70.Fh; 64.60.Cn; 75.10.Hk

\end{abstract}
\clearpage

\newpage
\section{Introduction}
More than 50 years ago C.N. Yang and T.D. Lee started a theory \cite{yl} for studying
phase transitions based on the partition function zeros . The same authors proved the
celebrated circle theorem in \cite{ly}. It assures, in particular, that the zeros of
the $s=1/2$ Ising model partition function on a given lattice, occur only at pure
imaginary values of the magnetic field. Defining $u=e^{-2\beta H} $ this implies that
the zeros are on the unit circle in the complex $u$ plane . The theorem is very robust
and does not depend on details of the lattice like space dimensionality, number of
nearest neighbors and topology. Later, the circle theorem was generalized to higher
spin models without \cite{griffths} and with\cite{suzuki73} extra symmetric
potentials, as well as, to ferroeletric models \cite{sf} and some Heisenberg models
\cite{sf,asano,dunlop}. In all those models the spin variables are defined on a given
static lattice and therefore they are the only relevant degrees of freedom. The proofs
are based on the fact that the partition functions can be written as polynomials of a
specific form \cite{ly}. Since addition of polynomials mix up their zeros in a rather
non trivial way it is unexpected that the circle theorem will hold for spin models
defined on dynamical random lattices which are themselves also degrees of freedom and
must be summed over altogether with the spins. However, numerical results for the
Ising model defined on 2D dynamical random lattices of planar
\cite{staudacher,ambjorn,fatfisher} and torus topology \cite{npb} indicate that the
Yang-Lee zeros do lie on the unit circle although no circle theorem is known in the
case of dynamical lattices. It is not even clear whether the new polynomials are of
the form assumed in \cite{ly}. In particular, the form of the polynomials used in
\cite{ly} are not kept by arbitrary linear combinations. The dynamical lattices used
in \cite{staudacher}-\cite{npb} were suggested in \cite{kazakov} in order to mimic the
effects of 2D fluctuating random surfaces in the continuum limit (2D gravity). The
partition functions were generated by $N\times N$ Hermitean random matrices and the 2D
gravity interpretation is achieved by fine tuning the coupling constants in the model
and taking $N\to\infty$ simultaneously. For the special case where the matrices become
numbers ($N=1$) \cite{bcp}, although the 2D gravity interpretation is lost, the models
defined on the so called thin graphs, exhibit phase transitions with mean field
exponents and strong similarities with models defined on the Bethe lattice
\cite{djbethe}. In this case the partition function is obtained by summing over spin
variables and connected and non-connected regular graphs (non-connected Feynman
diagrams). Once again, numerical results valid for finite number of spins \cite{npb}
and analytic results valid in the thermodynamic limit \cite{djedge,pre} strongly
support a circle theorem for this type of random lattices but no explicit proof for
finite $n$ has ever been given. It is the purpose of this work to produce such a proof
at least in the simplified case of one dimensional models where the random graphs
become connected and  non-connected rings (see fig. 1). We have chosen the
ferromagnetic Blume-Capel (BC) model \cite{bc} since it is the simplest model with
``would be" ${\cal Z}_2$ symmetry such that on one hand their zeros ,in any dimension,
are known to satisfy the circle theorem  in a certain parameters range \cite{suzuki73}
and on the other hand they have been shown in 2D to exhibit non trivial topological
properties like bifurcation for other range of the parameters \cite{biskupetal}.
Besides, the BC model includes the $s=1/2$ and $s=1$ Ising models.

\section{The 1D Blume-Capel model on non-connected rings}

In the spin one Ising model there are three spin states on each site : $S_i = -1,0,1
\, $. The Blume-Capel (BC) model is a generalization of the spin one Ising model
where, besides the magnetic field $H$ which works as a source for dipole interaction,
we now introduce a coupling $\lambda $ functioning as a source for quadrupole
interactions such that we still have an ``would be'' ${\cal Z}_2$ symmetry by
exchanging $H  \to -H$. The energy and the partition function of the Blume-Capel model
are given by:

 \be E\,=\,-\,J\,\sum_{\left<ij\right>}\,S_i\,S_j\,-\,H\,
\sum_{i=1}^n\,S_i\,-\,\lambda\,\sum_{i=1}^n\,(1\,-\,S_i^2) \label{ebc}\ee

\be Z_n\,=\,\sum_{\left\{S_i\right\}}\,e^{K\,\sum_{\left<ij\right>}\,
S_i\,S_j\,+\,\sum_{i=1}^n\,[h\,S_i\,+\,\Delta\,(1\,-\, S_i^2)\,]} \label{zbc}\ee

\no We only work with the ferromagnetic case $J > 0$. The sum $\sum_{\left<ij\right>}$
is over nearest neighbors sites. For the one dimensional model we assume periodic
boundary conditions $S_j=S_{n+j}$. In this case the lattice becomes a ring with $n$
sites (vertices) and $n$ bonds. Throughout this work we use the following definitions
:

\bea K\, =\, J/kT \quad &;& \quad c \, = \, e^{-K} \nn \\
h\, = \, H/kT &;& \quad u \, = \, e^{h} \label{constants}\\
\Delta \, =\, \lambda /kT \quad &;& \quad  x\, =\, e^{\Delta} \nn \eea

Although originally suggested in a magnetic context \cite{bc}, the BC model
corresponds to a special case of the BEG model \cite{beg} and it can be used to
describe phase separation driven by superfluidity in He$^{(3)}$-He$^{(4)}$ mixtures.
Its phase diagram in 2D is reach, including tricriticallity. At $x=0 \, (\Delta\to
-\infty )$ it becomes the $s=1/2$ Ising model while for $x=1 \, (\Delta=0)$ it
corresponds to the $s=1$ Ising model.

 Each configuration with $p$ spins down, $n_0$ spins zero, and $n_{ab}$ bonds
connecting spins $S_a$ to $S_b$ will contribute to the partition function a factor:

\be e^{\left\lbrack (n-p-n_0)h -p h + n_0 \Delta\right\rbrack } \, e^{\left\lbrack
K(n_{++} + n_{--}) - K n_{+-} \right\rbrack } \label{factor1} \ee

\no That factor will be  needed in order to define the model on Feynman diagrams
below. We start with the following generating function of  Feynman diagrams in a zero
dimensional field theory:

\bea G(h,c,\Delta,g) \, &=& \, \frac{\int\, d\phi_+\, d\phi_- d\phi_0 \, e^{-\frac 12
\left\lbrack \phi_a M_{ab}\phi_b - g \left( e^h \phi_+^2 + e^{-h} \phi_-^2 +
e^{\Delta} \phi_0^2\right)\right\rbrack  }}{ \int\, d\phi_+\, d\phi_- d\phi_0\,
e^{-\frac 12
\left\lbrack  \phi_a M_{ab}\phi_b \right\rbrack  }}\label{z2} \nn \\
\, &\equiv & \left\langle e^{\frac g2 \left( e^h \phi_+^2 + e^{-h} \phi_-^2 +
e^{\Delta} \phi_0^2 \right)} \right\rangle \label{ev1}\eea

\no where the coupling constant $g$ will play the role of a counting parameter while
the $3 \times 3$  symmetric matrix $M_{ab} , \, a,b=+,-,0$ is to be determined from
the Boltzmann weights in (\ref{zbc}) for $h=0=D$. The zero dimensional fields
$\phi_{\pm},\phi_0$ will represent the  spin states $S_i = \pm 1 , 0$ respectively.

\no The expansion :

\be G(h,c,g) \, = \, \sum_{n=0}^{\infty} g^n {\cal Z}_n^{nc} \quad ,
\label{expansion1} \ee

\no defines the coefficients

\be  {\cal Z}_n^{nc} \, = \, \frac 1{2^n n! } \left\langle \left( e^h \phi_+^2 +
e^{-h}\phi_-^2 + e^{\Delta} \phi_0^2\right)^n \right\rangle \label{znc} \ee

\no If we interpret $e^h \phi_+^2 $ , $e^{-h} \phi_-^2 $ and $e^{\Delta} \phi_0^2 $ as
interaction vertices representing sites with up, down and vanishing spins
respectively, the $ {\cal Z}_n^{nc} $ have an interpretation as a  diagrammatic
expansion {\it a la } Feynman. In figure 1 we draw them for $n=1,2,3$. The $ {\cal
Z}_n^{nc} $ correspond to the sum of all diagrams with a total of $n$ interaction
vertices such that each vertex has two lines attached and all lines must be connected
as in figure 1. Each vertex must be of course of type up, down or zero and will carry
a factor $e^h, e^{-h}$ and $e^{\Delta}$ respectively. Thus, we end up summing
 over all spin configurations altogether with all connected and non-connected rings with
$n$ vertices. The superscript ``nc'' stands for non-connected since the expansion
includes non-connected diagrams like the second one in the second row and the second
and third ones in the third row of fig.1. Each internal line of the diagram connecting
a vertex of type $a$ to a vertex of type $b$ corresponds mathematically to a Gaussian
integral $<\phi_a \phi_b > = M^{-1}_{ab}$ see (\ref{ev1}). Therefore, according to the
so called Feynman rules\footnote{For a demonstration of the Feynman rules in a simple
case see \cite{biz}}, each diagram with $p$ minus and  $n_0$ vanishing spin  contains
 a factor

\begin{figure}
\begin{center}
\epsfig{figure=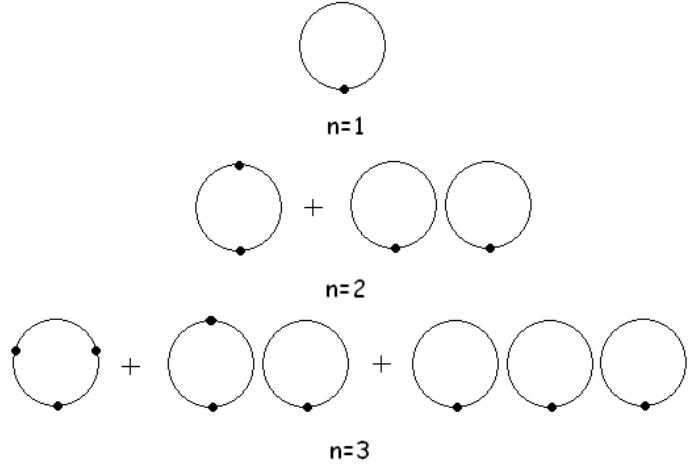,width=90mm} \caption{The Feynman diagrams corresponding to
${\cal Z}_1^{nc}$, ${\cal Z}_2^{nc}$ and ${\cal Z}_3^{nc}$.} \label{figure1}
\end{center}
\end{figure}

\be e^{\left\lbrack (n-p-n_0) h - p h + n_0 D \right\rbrack } \,
\prod_{a<b}\left\langle \phi_a\phi_b \right\rangle^{n_{ab}}\, \label{factor2}\ee

\no In order to make more contact with the statistical model and match the above
factor with (\ref{factor1}) we require  $ <\phi_a \phi_b > $ to be proportional to the
Boltzmann weights in (\ref{zbc}) calculated at $h=0=\Delta$. That is,

 \bea <\phi_a\phi_b> \, = \, {\bf M}^{-1}_{ab} \, &=& \,
 \kappa \, e^{-\frac{E_{ab}(h=0=\Delta)}{kT}}\, \nn \\ &=& \,
  \kappa \cdot \left(\begin{array}{ccc}
e^K & e^{-K} & 1\\
e^{-K} & e^K & 1\\
1 & 1 & 1\\
\end{array}\right)\nn  \eea

\no If we choose the overall constant $\kappa= c/\left\lbrack
(1-c^2)(1-c)\right\rbrack $ we obtain a simple form for $M_{ab}$ such that the
generating function  (\ref{ev1}) becomes

\be G(c,h,\Delta,g) \, =\, \frac{\int d\phi_+ \, d\phi_- \, d\phi_0 e^{-S_g}}{\int
d\phi_+ \, d\phi_- \, d\phi_0 e^{-S_{g=0}}} \label{bcexpansion}\ee

\no With

\bea S_g &=& \frac12\,\left\lbrack \,\phi_+^2+\phi_-^2+f
\phi_0^2+2\,c\,\phi_+\,\phi_--2\,(1+c)\,(\phi_+ + \phi_-)\,\phi_0\,\right\rbrack \nn
\\ &-&\frac{g}2\,(e^h \phi_+^2+e^{-h} \phi_-^2+e^{\Delta} \phi_0^2) \label{sg}\eea

\no and

\be f \, = \, \frac{(1+c)(1+c^2)}c \label{f} \ee

\no The Gaussian integrals in (\ref{bcexpansion}) can be easily calculated implying:

\be G(c,h,\Delta,g) \, =\,
\left\lbrack\frac{(1-c^2)^2(1-c)}{P_3(g)}\right\rbrack^{1/2} \label{exactg} \ee

\no Where

\be P_3(g)  \! = \! (1-c^2)^2(1-c)- g (1-c^2)(\tilde{x} + A) + g^2 \left\lbrack
(1+c)(1+c^2) + A \tilde{x} \right\rbrack - g^3 \tilde{x} \label{p3}\ee

\be A = e^h + e^{-h} \quad ; \quad \tilde{x} \, = \, x c \, = \, e^{\Delta -
K}\label{xt} \ee

\no Therefore, from (\ref{expansion1}) we can write down ${\cal Z}_n^{nc}$ as:

\be {\cal Z}_n^{nc} \, =\,
\left\lbrace\left\lbrack\frac{(1-c^2)^2(1-c)}{P_3(g)}\right\rbrack^{1/2}\right\rbrace_{g^n}
\label{zncdef1} \ee

\no or equivalently from (\ref{znc}):

\be {\cal Z}_n^{nc} \, =\, \frac{\int d\phi_+ \, d\phi_- \, d\phi_0 e^{-S_g}\left( e^h
\phi_+^2 + e^{-h}\phi_-^2 + e^{\Delta} \phi_0^2\right)^n}{2^n n! \,\int d\phi_+ \,
d\phi_- \, d\phi_0 e^{-S_{g=0}}} \label{zncdef2}\ee

\no It is instructive to relate ${\cal Z}_n^{nc}$ given in (\ref{zncdef1}) or (\ref
{zncdef2}) with the usual partition functions of the one dimensional BC model on a
ring ${\cal Z}_n$ given in (\ref{zbc}). The Feynman rules and our definition for the
overall constant $\kappa $ allow us to write the diagrammatic expansion:

\be {\cal Z}_n^{nc} \, = \, \left\lbrack\frac c{(1-c^2)(1-c)}\right\rbrack^n\,
\sum_{{\rm diagrams}}\frac{{\cal Z}_n(diagram)} {S_n({\rm diagram})} \label{znc1} \ee

\no Where the sum is over all diagrams with $n$ vertices, connected and non-connected
. The factor $S_n$ is a positive real number called symmetry factor of the graph and
it comes from the sum over all possible ways of connecting the vertices among each
other to build the corresponding graph. The number of such possibilities we call
$A_n$. According to our notation, see (\ref{znc}), and the Feynman rules $S_n= (2^n
n!)/A_n $. In order to obtain $A_n$ it is convenient to label the two lines of the
i-th vertex as $(a_i,b_i)$. For example, for the case of connected rings,like the
first diagrams on each row of fig. 1, there will be a total of $2n$ labels, if we
start with the label $a_1$ it can be linked with any of the labels $a_j$ or $b_j$ such
that $j\ne 1$. Thus, we have $2n-2$ possibilities for the first connection. The yet
non-connected label $b_1$ can be linked with  the labels of any vertex but the first
and the j$^{th}$ ones. Thus, there are $2(n-2)$ possibilities. If we keep going until
no label is left unconnected we end up with $A_n=(2n-2)!!$ and $S_n({\rm connected
\,\, ring}) \, = \, \frac{2^n n!}{(2n-2)!!} \, = \, 2 n $. The symmetry factor of the
non-connected diagrams can be similarly obtained. As an example, for $n=3$ (third row
in fig. 1) the expansion (\ref{znc1}) becomes:

\be {\cal Z}_3^{nc}\, = \, \left\lbrack \frac c{(1-c^2)(1-c)}
\right\rbrack^3\left\lbrack \frac{{\cal Z}_3}{6} \, + \, \frac{{\cal Z}_{1}{\cal
Z}_2}{8} \, + \, \frac{{\cal Z}_{1}^3}{48} \right\rbrack \label{znc3} \ee

\no The general expansion for arbitrary $n$ can be obtained as follows. First, it is
well known in field theory that the connected diagrams can be singled out by taking
the logarithm of the generating function. Using that $S_n = 2n$ for connected diagrams
we have:

\be \ln G(h,c,\Delta,g) \, = \, \sum_{n=1}^{\infty} \frac 1{2n}\left\lbrack \frac {g
\, c}{(1-c^2)(1-c)} \right\rbrack^n {\cal Z}_n \quad , \label{expansion2} \ee

\no Where on the right handed side we have only partition functions of connected
rings. Exponentiating (\ref{expansion2}) and comparing with the expansion
(\ref{expansion1}) we have the generalization of (\ref{znc3}) for arbitrary number of
vertices:

\bea {\cal Z}_n^{nc}(c,h) \, &=& \, \left\lbrack \frac
c{(1-c^2)(1-c)}\right\rbrack^n\left\lbrack \exp \left( \frac{g {\cal Z}_1}{2} +
\frac{g^2 {\cal Z}_2}{4} + \cdots + \frac{g^n {\cal Z}_n}{2n} \right)
\right\rbrack_{g^n} \\
&=& \left\lbrack \frac c{(1-c^2)(1-c)} \right\rbrack^n \left\lbrack \frac{{\cal
Z}_1^n}{2^n n! } + \cdots + \frac{{\cal Z}_1 {\cal Z}_{n-1}}{4(n-1) } + \frac{{\cal
Z}_n}{2n} \right\rbrack \label{zncn} \eea

\no In particular, compare with (\ref{znc1}),  the symmetry factors of the
non-connected diagrams are produced automatically. The expression (\ref{zncn}) shows
that a generalization of the  circle theorem valid for ${\cal Z}_n$ to  sums like
${\cal Z}_n^{nc}$ is non trivial.

\section{Yang-Lee Zeros}

\subsection{1D BC model on a connected ring}

Before we start with the 1D BC model defined on Feynman diagrams we first study the
zeros of the static lattice (one ring) model.

The partition function (\ref{zbc}) is a polynomial on each of the variables $c,x$ and
$u$. In particular, ${\cal Z}_n$ is a polynomial of degree $2n$ in the ``fugacity''
$u$. Concerning the Yang-Lee zeros on the complex $u$-plane, it was proven in
\cite{suzuki73} that they must satisfy the circle theorem for $0\le x \le 2$
independently of lattice details like topology, coordination number or space
dimensionality. The proof, reproduced below, is based on the decomposition
\cite{griffths} of a spin one particle in a cluster of two spin one-half ones: $S_i =
(\sigma_{i1} +\sigma_{i2})/2 $, where $\sigma_{i1} ; \sigma_{i2} = \pm 1$. After such
decomposition we can rewrite, up to an overall constant, ${\cal Z}_n$ as a partition
function of a bigger system with $2n$ spin one-half particles:

\be {\cal Z}_n \, \propto \, \tilde{\cal Z}_{2n} \, = \, \sum_{\{\sigma_{\alpha}\}}\,
\exp{\left\lbrack \sum_{< \alpha\beta
>}\tilde{K}_{\alpha\beta}\,\sigma_{\alpha}\sigma_{\beta} \, + \, \frac h2
\sum_{\alpha=1}^{2n} \sigma_{\alpha}\right\rbrack}\label{ztilde}\ee

\no where $\sigma_{\alpha}$ include all $\sigma_{i1}$ and $\sigma_{i2}$ with
$i=1,\cdots , n$. The sum $\sum_{< \alpha\beta >}$ runs over all nearest neighbors
(n.n.), including spins in the same cluster. The new coupling is:

\be \tilde{K}_{\alpha\beta} \, = \, \cases{\frac 12 \left(\ln 2 - \Delta \right) & \,
within \,  a \, cluster.  \cr K/4  & \, between \, n.n. \, of \, different\,
clusters.} \ee

\no The factor $\ln 2/2$ comes from the double degeneracy of the $S=0$ state which can
be represented inside a cluster by the weight $W(\sigma_{\alpha},\sigma_{\beta}) \, =
\, \exp\left\lbrack \frac 12 \ln2\left(\sigma_{\alpha}\sigma_{\beta}-1\right)
\right\rbrack$. Since the circle theorem holds for any ferromagnetic partition
function of the form (\ref{ztilde}) we conclude that for $\Delta \le \ln 2$ ($0\le x
\le 2 $), see \cite{suzuki73}, the Yang-Lee zeros of the BC model will lie on the unit
circle for any space dimension.

To the best we know, even though we will be dealing with the exactly solvable one
dimensional model, there are no exact analytic results about the Yang-Lee zeros of the
BC model for $\Delta > \ln 2$. The best we can do is to present what we found
numerically for finite number of spins and deduce approximate analytic results along
the lines of the work of \cite{glumac} for the one dimensional  Potts model. The
starting point for the numerical calculations of the  zeros is the exact expression
for ${\cal Z}_n$ which can be obtained , for instance, by inverting the expansion
(\ref{expansion2}) using (\ref{exactg}),

\bea {\cal Z}_n \, &=& \, 2n \left\lbrack\frac{(1-c^2)(1-c)}{c}\right\rbrack^n
\left\lbrace\ln
\left\lbrack\frac{(1-c^2)^2(1-c)}{P_3(g)}\right\rbrack^{1/2}\right\rbrace_{g^n} \label{exactzn1}\\
&=&\, \left(\lambda_+^n + \lambda_-^n + \lambda_0^n\right)/c^n \label{exactzn2}\eea

\no where $\lambda_{\pm}$ and $\lambda_0$ are solutions of the cubic equation:

\be \lambda^3 - \lambda^2 \left( A + \tilde{x} \right) + \lambda \left\lbrack A
(1-c)\tilde{x} + 1- c^4\right\rbrack - \tilde{x} (1-c^2)(1-c)^2 \, = \, 0
\label{cubic} \ee

Incidentally, we notice that the inversion of formula (\ref{expansion2}) is a
diagrammatic method alternative to the transfer matrix approach. The quantities
$\lambda_{\pm}/c \, ; \, \lambda_0/c$ are precisely the eigenvalues of the transfer
matrix of the 1D BC model, although we never really used it in our derivation of
(\ref{exactzn2}) which came from factorizing $P_3(g)$.  We remark that the way we
found  the exact solution (\ref{exactzn2}) is a combinatorial one, where the Feynman
rules take care of the correct combinatorial factors. In this sense it is similar to
the original solution of the one dimensional Ising model by E.Ising \cite{ising}.

Proceeding further, for the purpose of generating the exact partition functions for
large $n$ it  turned out to be faster to use the expression (\ref{exactzn1}) instead
of plugging, for fixed $c$ and $\tilde{x}$,  the solution of (\ref{cubic}) in
(\ref{exactzn2}). For other higher spin models $s > 1$ the computational advantage of
(\ref{exactzn1}) against (\ref{exactzn2}) is even bigger. Given the exact expressions
for the partition function, we have numerically obtained their zeros using the
software Mathematica. Among other checks we have used the exact analytic results for
the zeros of the 1D Ising model on one ring  to confirm the accuracy of the zeros. As
a general feature we have found numerically that the complex $u=e^h$ zeros never
converge for non-zero temperatures to the positive real axis in agreement with the
fact that no phase transition for real magnetic fields appears in one dimensional spin
models with short range interaction. Furthermore, the zeros lie on the unit circle for
$0\le x \le 2 $ and arbitrary temperatures $0 \le c \le 1$, thus confirming the
analytic result of \cite{suzuki73}. For any given $x>2$ we have always been able to
find a temperature after which the zeros departure from the unit circle as shown in
figures 2a-2d.

\begin{figure}
\begin{center}
\epsfig{figure=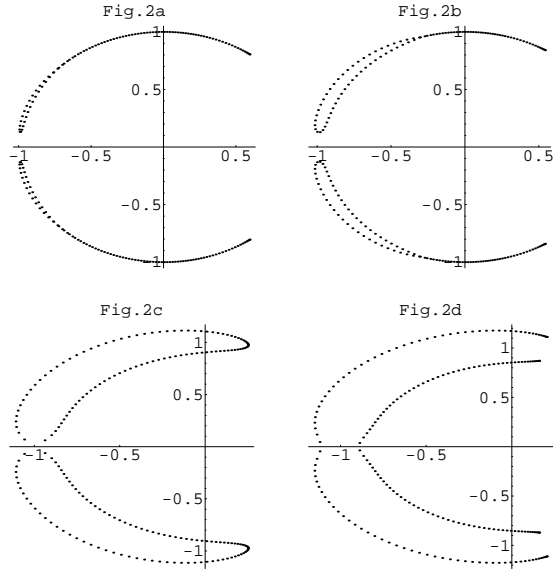,width=80mm} \caption{Yang-Lee zeros of the usual 1D
Blume-Capel model on the complex $e^h$ plane  for $x = 3$ and temperatures $c=0.435 ;
0.450 ; 0.500 $ and $ 0.520$ respectively in figures $2a , 2b , 2c$ and $2d$.}
\label{figure2}
\end{center}
\end{figure}

\no As $x$ increases the minimal temperature for which that happens decreases. As a
general rule, increasing $x$ or $c$ or even both of them the zeros move in a such way
that after a given value for the product $x c = e^{\Delta -K} $ they start bifurcating
into two branches: one inside and the other outside the unit circle. A bifurcation of
the Yang-Lee zeros has been also observed in the two dimensional BC model
\cite{biskupetal} but in that case the bifurcation starts at $\theta =0$ instead of
$\theta = \pi$. Our numerical results show that in the interval $0\le \tilde{x} \le 1$
the zeros lie on the unit circle\footnote{The numerical results presented in
\cite{biskupetal} at a fixed temperature in the 2D case are in approximate agreement
with the interval $0 \le \tilde{x} \le 1$, where $e^{-\lambda}$ of that reference
corresponds to  $\tilde{x} = c x$, see comment below figure 1 of that reference.}
Although we have not been able to deduce a rigorous proof of this fact analytically,
we have approximate analytic results based on the approach of \cite{glumac}. The main
idea is that the zeros of (\ref{exactzn2}) appear whenever at least two transfer
matrix eigenvalues have the same absolute value which on its turn must be bigger than
the absolute value of the remaining eigenvalue, e.g., $\vert \lambda_+ \vert = \vert
\lambda_- \vert > \vert \lambda_0 \vert$. Unfortunately, differently from the Q-state
Potts model treated in \cite{glumac} which is similar to the Ising model ($Q=2$), the
secular equation (\ref{cubic}) is not easily factorizable and their explicit solutions
are not very useful to match absolute values. We have to stick to approximate
solutions. In order to keep some similarity to the Ising model ($x=0$), we solve
(\ref{cubic}) as a series expansion for low $\tilde{x}$. That gives:

\bea \lambda_0 \, &=& \, \frac{(1-c^2)^2 \tilde{x}}{(1+ c^2)} - \frac{2 A c (1-c)^2
\tilde{x}^2 }{(1+c^2)^3} + {\cal O}(\tilde{x}^3) \label{l0} \\
\lambda_{\pm} \, &=& \, \sqrt{1-c^4} \, g_{\pm} + \left\lbrack c +
\frac{2c^3}{(1+c^2)(g_{\pm}^2 -1)} \right\rbrack \, \tilde{x} + {\cal O}(\tilde{x}^2)
\label{lpm}\eea

\no where

\be g_{\pm} \, = \, \frac A{\sqrt{1-c^4}} \, \pm \, \left\lbrack \left(\frac
A{\sqrt{1-c^4}}\right)^2 - 1 \right\rbrack^{1/2} \label{gpm} \ee

\no The singularity at $g_{\pm}^2=1$ is the Yang-Lee edge singularity ($\lambda_+ =
\lambda_- $) of the Ising model ($\tilde{x}=0$). Since $g_+g_-=1$ we can write $g_+ =
\rho e^{i\theta}$ and $g_- = e^{-i\theta}/\rho $. Therefore, we write $\vert \lambda_+
\vert^2 = r(\rho,\theta,c,\tilde{x}) r(\rho,-\theta,c,\tilde{x})$ and $\vert \lambda_-
\vert^2 = r(1/\rho,-\theta,c,\tilde{x}) r(1/\rho,\theta,c,\tilde{x})$ for some
function $r(a,b,c,d)$. Thus, assuming  $\rho=1$ ($A = \cosh (h) = \sqrt{1-c^4}$),
which leads us to the unit circle $\vert e^h \vert =1$ since $0\le \cosh (h) \le 1$,
we have $\vert \lambda_+ \vert = \vert \lambda_- \vert $. Therefore, our assumption
will lead in fact to zeros on the unit circle if  $\delta_{\pm} = \vert \lambda_{\pm}
\vert - \vert \lambda_0 \vert > 0 $, or equivalently $\delta_{\pm} = \Re e(f_{\pm} -
f_0) > 0 $ where $f_i =  \ln\lambda_i$. This can be checked from (\ref{l0}) and
(\ref{lpm}). Using our assumption $g_{\pm} = e^{\pm i\theta} $ we get :

\be \delta_{\pm} \, = \, \frac 12 \ln\left\lbrack
\frac{(1+c^2)^3(1+c)}{(1-c)^3}\right\rbrack \, - \, \ln \tilde x \, + \tilde{x}\, c
\left\lbrack\frac{5 \cos\alpha}{(1+c^2)^2} \right\rbrack\, + \, {\cal
O}\left(\tilde{x}^2\right) \label{deltapm} \ee

\no Where $h\equiv i\alpha$, consequently $\cos\theta= \cos\alpha/\sqrt{1-c^4}$. The
first term on the right handed side of (\ref{deltapm}) is always positive and the
second one shows why the zeros go to the unit circle for $0 \le \tilde{x} \le 1 $. The
third term is also in agreement  with our numerical calculations since it is more
negative, thus pushing the zeros outside the unit circle, for angles close to $\alpha
= \pi$. Adding up the three terms we have checked that $\delta_{\pm} > 0$ whenever
$\tilde{x} \le \tilde{x}^* $ where the minimum value for $\tilde{x}^*_{min} =
0.780(2)$ occurs at $c=0.220(5)$ and $\alpha = \pi$. The fact that
$\tilde{x}^*_{min}\ne 1$ is an artefact of the low $\tilde{x}$ approximation. In
practice our numerical calculation of the zeros show that they lie on the unit circle
for $0 \le \tilde{x} \le 1 $ with great accuracy. Although that seems to be a weak
form (temperature dependent) of the circle theorem, one can slightly redefine the BC
model, as in \cite{biskupetal}, such that the three configurations where {\it all}
spins are on the same state become the zero temperature degenerate ground states. This
redefinition ($x\to x c = \tilde{x}$) makes $\tilde{x}$ temperature independent,
however the results derived  so far for the zeros on the complex $u=e^h$ plane are
still the same.

After this partial understanding of the conditions under which the Yang-Lee zeros
remain on the unit circle, one might ask what can we say  about the parameters region
for which the Lee-Yang circle theorem is not valid. We do not have a nonperturbative
approach to this question but we have made low temperature expansions (LTE) analogous
to (\ref{deltapm}). First of all, we found useful to implement the redefinition $x\to
x c = \tilde{x}$ before taking $c\to 0$. We emphasize that such redefinition does not
change the position of the zeros . Solving (\ref{cubic}) by a power series on $c$ we
have:

\bea \lambda_0 &=& \xt  \, + \, \frac{2(A \xt -1)\xt}{(\xt - e^h)(\xt -
e^{-h})}c \, + \, {\cal O}\left(c^2\right)\label{l0b}\\
\lambda_{\pm} &=& e^{\pm h} \left( 1 \, + \, \frac{\xt}{e^{\pm h} - \xt}c \, \right)+
\, {\cal O}\left(c^2\right) \label{lpmb} \eea

\no Defining $ e^h = \rho e^{i\alpha} $, imposing that $\Re e (f_+ - f_0)=0$ or $\Re e
(f_- - f_0)=0$ we obtain respectively:

\bea \rho = \rho_+  &=&  \xt \, + \, \frac{\xt\left(\xt^2\cos\alpha -1\right)}{1+
\xt^4
-2\xt^2\cos\alpha} c \, + \, {\cal O}\left(c^2\right)\nn \label{rhop} \\
\rho = \rho_-  &=&  \frac 1{\xt} \, - \, \frac{\left(\xt^2\cos\alpha -1\right)}{\xt
(1+ \xt^4 -2\xt^2\cos\alpha )} c \, + \, {\cal O}\left(c^2\right)\nn \label{rhom}\eea

\no Notice that, due to the $H \to - H$ symmetry which exchanges $f_+$ and $f_-$ we
have $\rho_+ = 1/\rho_- $. Assuming $\vert\lambda_+\vert = \vert\lambda_0\vert$ we are
led to $\rho=\rho_+$, in this case we have calculated the difference:

\be \Re e (f_+ - f_-) \, = \, 2\ln\xt + \frac{3 \left(\xt^2\cos\alpha -1\right)}{\xt^4
+1-2\xt^2\cos\alpha} c \, + \, {\cal O}\left(c^2\right) \equiv F(\xt,c,\alpha) \quad .
\label{dpm}\ee

\no  Whenever $F > 0$ we have $\vert\lambda_+\vert = \vert\lambda_0\vert >
\vert\lambda_-\vert $. If  we had assumed $\rho=\rho_-$ in calculating  $\Re e (f_+ -
f_-)$ we would simply exchange the sign of the result, i.e., $ \Re e (f_+ - f_-) = -
F$ and consequently for $F > 0$  we would have $\vert\lambda_-\vert =
\vert\lambda_0\vert > \vert\lambda_+\vert $. Concluding, for $F > 0$ the zeros may
come
 either from $\vert\lambda_+\vert = \vert\lambda_0\vert > \vert\lambda_+\vert$
with $ \rho =\rho_+$ or
 $\vert\lambda_-\vert = \vert\lambda_0\vert > \vert\lambda_+\vert$
 with $ \rho =\rho_-$ while for $F < 0 $
 they come from $\vert\lambda_+\vert = \vert\lambda_-\vert > \vert\lambda_0\vert$
 with $\rho =1$. Therefore,
 the form of the function $F$ shows, at least for low temperatures, why the zeros departure
 from the unit circle as $\xt $ increases beyond $\xt =1$. In figure 3 we plot $\rho_+$ and
 $\rho_-$ for
$c=0.1$ and $\xt=1 ; 1.1 ; 1.2$ altogether with the numerical results for the zeros
for $n=100$ spins.

\begin{figure}
\begin{center}
\epsfig{figure=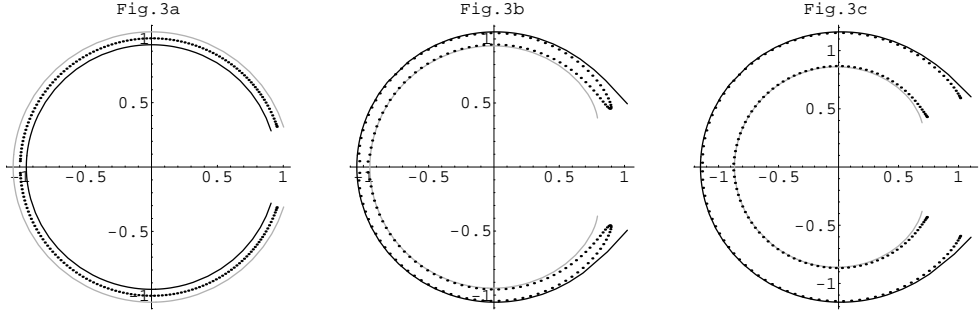,width=140mm} \caption{Overlap plot of $\rho_+$ (dark
solid),$\rho_-$ (light solid) and Yang-Lee zeros (dots) of the usual 1D Blume-Capel
model at $c=0.1$ and $\tilde{x}= 1.0 \,$(fig.3a) $ \, , 1.1 \, $(fig.3b) and $ 1.2\, $
(fig.3c). In fig. 3a the zeros lie on the unit circle. In all figures we used $n=100$
spins (200 zeros). } \label{figure3}
\end{center}
\end{figure}

We see that the zeros inside (outside) the unit circle basically come form
$\vert\lambda_-\vert = \vert\lambda_0\vert $ ( $\vert\lambda_+\vert =
\vert\lambda_0\vert $ ) while the zeros on the unit circle originate  from
$\vert\lambda_+\vert = \vert\lambda_-\vert $. One can check that $F>0$ for the cases
of figure 3b and 3c but this is not true for figure 3a which explains the disagreement
with the analytic curves. The large deviations which appear in the region close to the
edges of the curves of zeros are consequences  of the bad behavior of the LTE around
the Yang-lee edge singularity. The same problem occurs in the LTE of the simpler case
of the $s=1/2$ 1D Ising model. In passing, we notice that the Yang-Lee edge
singularities can be obtained by imposing the degeneracy of the two largest
eigenvalues of the  transfer matrix as in \cite{wang}. We have used those
singularities as a double check on the position of the zeros at the edges of the arcs.
For the number of spins used in figures 3 the Yang-Lee singularities basically overlap
with such zeros. In particular, in figure 3c there are four Yang-Lee edge
singularities coinciding with the four edges of the two arcs. Finally, the reader may
ask whether a triple degeneracy is possible, i.e., $\vert \lambda_+ \vert = \vert
\lambda_- \vert = \vert \lambda_0 \vert $. In practice we never found this type of
zero in our numerical results. After an exact analysis of the secular equation
(\ref{cubic}) we have found a necessary but not sufficient condition :  $\tilde{x} >
\sqrt{1+c} $.

\subsection{1D BC model on Feynman diagrams}

We begin with the simplest case of the 1D Ising model on Feynman diagrams ($x=0$).
Only in this special case we have $\lambda_0=0$ and the polynomial $p_3(g)$ becomes a
second order polynomial such that the  ${\cal Z}_n^{nc}$, as given in (\ref{zncdef1}),
turn out to be proportional to Legendre polynomials (${\cal P}_n$) as follows:

\bea  {\cal Z}_n^{nc} \, &=& \, \left\lbrack \frac{1}{\sqrt{\tilde{g}^2 - 2
\tilde{g}\,\frac{\cosh (h)}{\sqrt{1-c^4}} + 1} } \right\rbrack_{g^n}
\label{G} \nn\\
&=& \,\left\lbrack \sum_{n=0}^{\infty} \tilde{g}^n \, {\cal P}_n \left(\frac{\cosh
(h)}{\sqrt{1-c^4}}\right)\right\rbrack_{g^n} \nn \\
&=& \, \left\lbrack \frac{1 + c^2}{(1-c^2)(1-c)^2}\right\rbrack^{n/2}\, {\cal P}_n
\left(\frac{\cosh (h)}{\sqrt{1-c^4}}\right) \quad , \label{legendre}\eea

\no where $\tilde{g} = g\, \sqrt{1+c^2}/\left\lbrack
(1-c^2)(1-c)^2\right\rbrack^{1/2}$.

It is known that polynomials which are orthogonal in the interval $\left\lbrack
a,b\right\rbrack $ must have their zeros inside such interval excluding the borders.
Therefore, the zeros of  the Legendre polynomials  ${\cal P}_n(y_i)=0 \, ; \,
i=1,\cdots , n$ are such that $-1 < y_i < 1$. The Yang-Lee zeros of the 1D Ising model
on Feynman diagrams can thus, be obtained from the zeros of the Legendre polynomials:
$\cosh (h_j)=\sqrt{1-c^4} \, y_j $ which implies that the magnetic field is pure
imaginary ($h_j=i \alpha_j $). So the Yang-Lee zeros must be on the unit circle on the
complex $u=e^h$ plane. This is so far the only exact proof of a circle theorem for
statistical models defined on Feynman diagrams valid for finite number of spins. For
two dimensions we found an analytic proof for the Ising model on Feynman diagrams with
cubic vertices in \cite{npb} but valid only in the thermodynamic limit and inside the
convergence region of the low temperature expansion.

In order to single out the effect of the Feynman diagrams on the zeros, it is
interesting to compare the zeros that we have just found with the ones of the usual 1D
Ising model with periodic boundary conditions (connected ring). We first notice that
each zero of the Legendre polynomial furnishes two Yang-Lee zeros $\pm\alpha_j^{nc}$
given by

\be \cos \alpha_j^{nc} \, = \, \sqrt{1-c^4}\, y_j \label{alphaj1} \quad . \ee

\no For the usual 1D Ising model ($x = 0$) we have $\lambda_0 =0$, see (\ref{cubic}),
and ${\cal Z}_n = (\lambda_+^n + \lambda_-^n)/c^n $. The Yang-Lee zeros come  from
$\lambda_{\pm} = r e^{\pm i \theta_k} $ where $\theta_k=(2k-1)\pi/(2 n) $ with
$k=1,2,\cdots , n$. From (\ref{cubic}) we have  $\lambda_+ + \lambda_- = A = 2 \cosh
(h)$ and $\lambda_+\lambda_- = (1-c^4)$. Therefore $r=\sqrt{1-c^4}$ and the zeros must
be on the unit circle $h_j=i\alpha_j $ with $\pm\alpha_j$ given by

\be \cos \alpha_j \, = \, \sqrt{1-c^4} \cos\frac{(2j-1)\pi}{2 n} \label{alphaj2} \quad
. \ee

Comparing (\ref{alphaj1}) with (\ref{alphaj2}) we see that, while the zeros of the
model on the static lattice could be analytically determined,  the sum over Feynman
diagrams have led us to more complicated zeros, though still on the unit circle. Only
numerical results can be obtained in general for zeros of Legendre polynomials. In the
thermodynamic limit $n\to\infty$ we have asymptotic formulae for the zeros of Legendre
polynomials, see e.g. \cite{benderorzag}, at leading order : $y_k = \cos\left\lbrack
(4k-1)\pi/(4n+2)\right\rbrack $ with $k=1,2,\cdots ,n$. Using this formula it is easy
to check that $\alpha_j - \alpha_j^{nc}$ is positive (negative) for zeros close to the
positive (negative) real axis and it approaches zero at $\alpha $ close to $\pi /2$ in
agreement with figure 4.

\begin{figure}
\begin{center}
\epsfig{figure=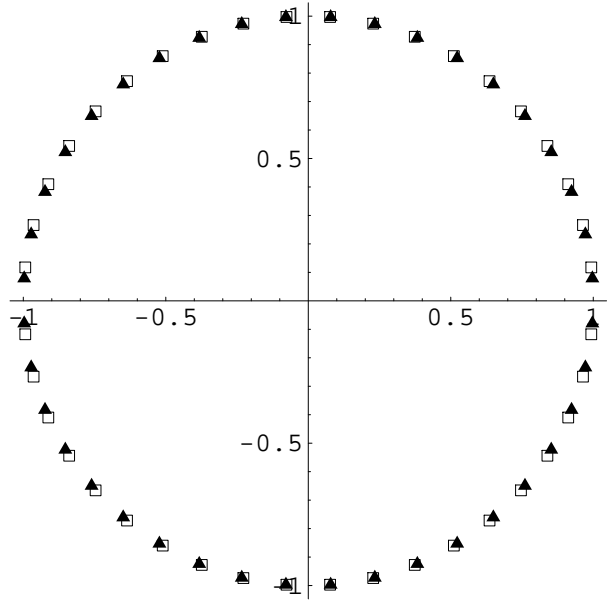,width=80mm} \caption{Yang-Lee zeros for the Ising model on
a connected ring (empty squares) and on Feynman diagrams (filled triangles) at $c=0.1$
and $n=20$ spins. } \label{figure4}
\end{center}
\end{figure}

Besides, the difference $\alpha_j - \alpha_j^{nc}$ is of order $ 1/n^2 $ in general
which means that in the thermodynamic limit the zeros of the  1D Ising model on
Feynman diagrams approach those of the model on the static lattice (one ring). At
$n=70$ spins we found already no visible difference between the zeros. This seems to
be a general feature of the 1D spin models defined on Feynman diagrams as we will see
later. It is remarkable that the zeros of the usual 1D Ising model with periodic
boundary conditions are directly related, in the thermodynamic limit, with the zeros
of Legendre polynomials.

Next we analyze  the general case of the 1D BC model on Feynman diagrams for arbitrary
$x$. Unfortunately, we have not been able to rewrite ${\cal Z}_n^{nc}$ in terms of
orthogonal polynomials for $x \ne 0$. Thus, for finite number of spins we had to find
the Yang-Lee zeros numerically.  It turned out faster to calculate the Gaussian
integrals in the  expression (\ref{zncdef2}) than to use the formula (\ref{zncdef1}).
We found the zeros with great accuracy for diagrams with up to $n=120$ spins, i. e.,
up to 240 zeros. We have checked for different temperatures,  different values of
$x=e^{\Delta}$ and various number of spins that whenever the circle theorem holds for
the usual 1D BC model it will also hold for the corresponding model on Feynman
diagrams. So, there must be a generalization of the circle theorem for Feynman
diagrams which is valid for finite size diagrams.

\begin{figure}
\begin{center}
\epsfig{figure=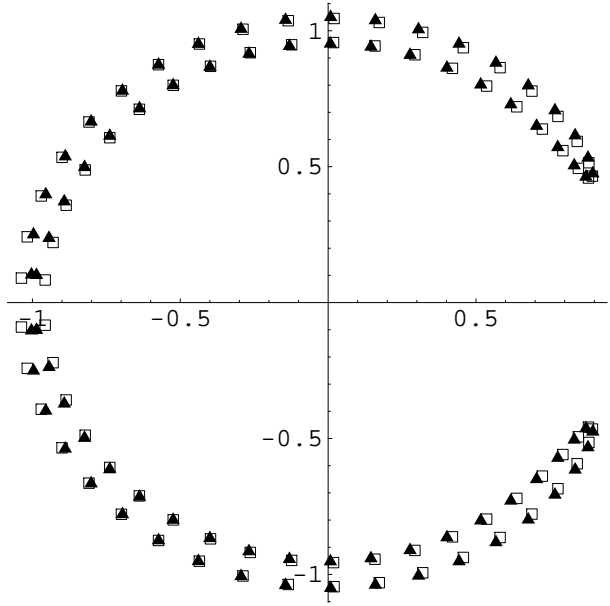,width=80mm} \caption{Yang-Lee zeros for the Blume-Capel
model on a connected ring (empty squares) and on Feynman diagrams (filled triangles)
at $c=0.1$ , $\tilde{x} = 1.1$ and $n=40$ spins. } \label{figure5}
\end{center}
\end{figure}

 Furthermore, as we approach
the thermodynamic limit we found in all cases that the Yang-Lee zeros on Feynman
diagrams always tend to the zeros on the static lattice (one ring), see figure 5 where
the zeros of the two models are already close for $n=40$ spins. This has been
confirmed by an argument based on the  saddle point equations as follows. After a
redefiniton $\phi_a \to \sqrt{n} \phi_a$ in the numerator of (\ref{zncdef2}) and
calculating the denominator we obtain:

\be {\cal Z}_n^{nc} \, = \, \frac{n^{n +3/2}(1-c^2)\sqrt{1-c}}{2^n n ! (\pi/2)^{3/2}
\sqrt{c} } \, \int d \phi_+ d\phi_- d\phi_0 e^{- \, n \, S} \label{zncsp1} \ee

\no Where $S = S_{g=0} - \ln V $ with $V= e^h \phi_+^2+e^{-h} \phi_-^2+e^{\Delta}
\phi_0^2 $. The saddle point equations defined by $\partial_a S =0 $ are given by:

\bea \phi_+ - (1+c) \phi_0 + c \phi_- \, &=& \, \frac{2 e^h \phi_+}{V} \nn\\
\phi_- - (1+c) \phi_0 + c \phi_+ \, &=& \, \frac{2 e^{-h} \phi_-}{V} \label{sp}\\
f \phi_0 - (1+c)\left( \phi_+ + \phi_-\right)  \, &=& \, \frac{2 e^{\Delta}\phi_0}{V}
\nn\eea

\no It is easy to combine the above equations to show that at all saddle points (s.p.)
solutions we have $S_{g=0} = 1 $ and $V$ satisfies a cubic equation such that
$(1-c^2)(1-c)V/2 \equiv \lambda $  satisfies precisely (\ref{cubic}). Thus, the six
solutions of the saddle point equations are doubly degenerated since they produce only
three different values for $S$ at leading order. Expanding about the saddle points
$\phi_a = \tilde{\phi}_a + \zeta_a/\sqrt{n} $ we have up to quadratic terms $S =
\tilde{S} + (\zeta_a\zeta_b
\partial_a \partial_b S\vert_{s.p.}) /2n $. Integrating over the fluctuations
$d\zeta_a$ and using Stirling's approximation we have at leading order:

\be {\cal Z}_n^{nc} \, = \, \frac{2(1-c^2)\sqrt{1-c}}{\sqrt{2\pi n c}}\left\lbrack
\frac 1{(1-c^2)(1-c)}\right\rbrack^n \, \sum_{s.p.}
\frac{\lambda_{s.p.}^n}{\sqrt{\det\partial_a
\partial_b S\vert_{s.p.}}} \label{zncsp2} \ee

\no where the factor two in the numerator comes from the double degeneracy of the
saddle point solutions and $\lambda_{s.p.} $ can only be $\lambda_+,\lambda_-$ or
$\lambda_0$. If we compare the symmetry factors in (\ref{zncn}) we notice that the
more non-connected diagrams like the first one between brackets drop out at
$n\to\infty$. However, at this limit, see (\ref{zncn}), the connected diagram is not
the only one to survive. If we remember the solution of the usual connected partition
function (\ref{exactzn2}) and look at (\ref{zncsp2}) we realize that the effect of the
other partition functions of the same order as ${\cal Z}_n $ is to produce the
denominators $\sqrt{\det\partial_a\partial_b S\vert_{s.p.}}$.

The sum in (\ref{zncsp2}) extends over all relevant saddle points. In order to single
out which solutions of (\ref{sp}) are relevant, a careful analysis of the absolute
values of $\lambda_a$ would be necessary which depends in general on the parameters
$c,e^{\Delta},e^{h}$ and is rather complicated if we want an exact result. Instead, we
discuss the case of the Ising model ($x=0$) and assume that the general case will be
similar as our numerical results for finite number of spins seem to indicate. For the
Ising model we know that ${\cal Z}_n^{nc}$ are proportional to Legendre polynomials
for which asymptotic formulas are known, see e.g. \cite{benderorzag} so we can check
the result. If $x=0$ we have $\lambda_0=0$. We are thus left with two doubly
degenerated saddle point solutions. We assume that they all contribute whenever
$\vert\lambda_+\vert = \vert\lambda_-\vert$ which means $\lambda_{\pm } =
\sqrt{1-c^4}\, e^{\pm i\theta}$. In this case (\ref{zncsp2}) becomes,

\bea {\cal Z}_n^{nc}  &=&  \frac{2 \sqrt{1-c^4}}{\sqrt{4 \pi n}(1-c^2)^n(1-c)^n}\,
\left\lbrace \frac{\lambda_+^n}{\left\lbrack\lambda_-\left(\lambda_+ -
\lambda_-\right)\right\rbrack^{1/2} }\, + \,
\frac{\lambda_-^n}{\left\lbrack\lambda_+\left(\lambda_- -
\lambda_+\right)\right\rbrack^{1/2}}\right\rbrace \label{zncising} \\
& = & \left\lbrack \frac {1+c^2}{(1-c^2)(1-c)^2}\right\rbrack^{n/2} \sqrt{\frac 2{\pi
n}} \, \frac{\sin \left\lbrack \left(n+\frac 12 \right)\theta + \frac{\pi}4
\right\rbrack}{(2 \sin\theta)^{1/2}} \label{zncising2} \eea

\no The above expression reproduces exactly, including the overall constant asymptotic
formulas for Legendre polynomials at leading order \cite{benderorzag}, compare with
(\ref{legendre}), which shows {\it a posteriori} that all four saddle point solutions
( doubly degenerated ) must contribute. The Yang-Lee zeros correspond to $\cos\alpha =
\sqrt{1-c^4} \cos\left\lbrack (4k-1)\pi/(4n-2) \right\rbrack $ with $k=0, \cdots , n$
as we mentioned before. Since $\lambda_{\pm}$ are the solutions of the equation
(\ref{cubic}) with $\xt =0$, i.e., $\lambda_{\pm} =  \cosh (h) \pm \sqrt{\cosh^2 + c^4
-1 } $. If $h$ and $c$ are real both $\lambda_{\pm}$ are also real and $\lambda_+
> \lambda_-$. In this case $\det\partial_a \partial_b S < 0 $ at the second
solution, see second term in (\ref{zncising}). Thus, the steepest descent trajectory
only passes by the first saddle point, the second  point  is not a true saddle point.
Once again we reproduce asymptotic formulas for ${\cal P}_n (y)$ for $y>1$. In
particular, the free energy of the model on Feynman diagrams is the same one of the
traditional 1D Ising model on a ring up to the addition of a constant.

 As a final comment concerning the 1D model Ising on Feynman diagrams,
we mention that the condition $\det \partial_a\partial_b S =0 $ at the saddle points
gives rise presumably \cite{djedge} to continuous phase transitions. In our case this
implies that the Yang-Lee edge singularity of the Ising model defined on Feynman
diagrams coincides with the same singularity in the usual 1D Ising model, i.e.,
$\lambda_+ = \lambda_- $.

For the general case of the 1D BC model on Feynman diagrams, once again, the zeros can
only appear whenever the absolute value of, at least, two eigenvalues $\lambda_a$
degenerate, which is the same basic equation for the zeros that we have obtained when
discussing the traditional 1D BC model on a ring. This equation gives $\rho=\rho
(\alpha)$ in the complex $e^h = \rho e^{i\alpha} $ plane. The precise values for
$\alpha $ will be fixed by the relative phases of the saddle points which are present
in the factors $\sqrt{\det\partial_a \partial_b S\vert_{s.p.}}$. According to our
numerical results for finite number of spins we believe that the $\alpha_k$ will be
close to their corresponding counterparts in the connected (traditional) 1D BC model.

\subsection{Conclusion}

\no Concerning the usual 1D Blume-Capel model defined on a connected ring (periodic
boundary conditions) our numerical results have shown that their Yang-Lee zeros remain
on the unit circle as far as either $\Delta \le \ln 2 $ or $ \Delta \le K $. The first
condition has been known for many years \cite{suzuki73} to be sufficient, in arbitrary
dimensions, to rewrite the Blume-Capel model in a such way that the original Lee-Yang
circle theorem applies. The second one is new and it is not clear whether it only
works in the one dimensional case. We have presented a perturbative proof of this
second condition valid for small $e^{\Delta - K }$. Furthermore, we have observed that
as a general feature as we increase $\Delta $ or the temperature or even both of them,
the zeros which lie originally on the unit circle tend to bifurcate as in the two
dimensional case \cite{biskupetal}, although no phase transition is present for real
magnetic fields and non-vanishing temperature in one dimension. At some point
$(\Delta^*,K^*)$ two disjoint arcs appear. We have shown by a low temperature
expansion that the existence of two curves can be understood from the degeneracy of
the absolute value of the transfer matrix eigenvalues. One of the curves comes from
$\vert\lambda_+\vert = \vert\lambda_0\vert $ while the other one is due to
$\vert\lambda_-\vert = \vert\lambda_0\vert $. For the zeros which remain on the unit
circle the degeneracy condition is: $\vert\lambda_+\vert = \vert\lambda_-\vert $.

Regarding the 1D Blume-Capel model defined on Feynman diagrams, which includes a sum
over connected and non-connected rings, although its partition function is rather
complicated, see (\ref{zncn}), its zeros seem to follow, specially for large rings,
the same pattern of the corresponding model defined on a connected ring. In
particular, the two previously mentioned conditions are apparently satisfied. We have
not been able to furnish a rigorous proof of this fact in general but an argument
based on the saddle point solutions show that the conditions for Yang-Lee zeros are
the same degeneracy conditions of the last paragraph. We believe that this is a
general feature of models defined on the type of dynamical lattice used here. This
indicates that a generalization of the circle theorem for dynamical lattices might
exist. In fact, in the special case of the 1D Ising model $\Delta \to - \infty $ we
have proved that the Yang-Lee zeros of the model defined on Feynman diagrams do lie on
the unit circle. This is the only rigorous proof we know valid for dynamical lattices
and finite number of degrees of freedom. It is not clear whether the polynomials can
be written in the form used in the original proof of the circle theorem in \cite{ly}
or whether  this proof is a independent one. Remarkably, it turns out that the zeros
in the variable $e^h$ of the partition function of the usual 1D Ising model with
periodic boundary conditions are directly related in the thermodynamic limit with the
zeros of the Legendre polynomials.

Finally, we remark that throughout this paper we have obtained the partition function
of usual 1D models with periodic boundary conditions (one ring) by taking the
logarithm of the generating function of all diagrams including non-connected ones.
This well known result in diagrammatic expansions furnishes an alternative to the
usual abstract but simpler transfer matrix approach in solving exactly 1D models. It
has already proven to be useful from the computational point of view since it became
faster than using solutions of the secular equation and plugging back in the partition
function.

\section{Acknowledgements}

This work was supported by CNPq, CAPES-Proap and FINEP (cc@complex). D.D. acknowledges
a discussion on connected diagrams with Luis C. de Albuquerque and an e-mail exchange
with Nelson. A. Alves.

\end{document}